\documentclass[12pt]{article}
\renewcommand{\baselinestretch}{1.2}
\usepackage{indentfirst}
\usepackage{amsmath}
\usepackage{amssymb}
\usepackage{amsfonts}
\usepackage{amscd}
\usepackage{amsbsy}
\usepackage{amsthm}
\usepackage{latexsym}
\usepackage{graphicx,color} %,epsf}
\usepackage[dvipsnames]{xcolor}
\usepackage[colorlinks]{hyperref}
\hypersetup{linkcolor=blue,citecolor=blue,urlcolor=blue}
\def\nn{\nonumber}       %%%    nonumber
\def\beq{\begin{eqnarray}}
\def\eeq{\end{eqnarray}}
\def\str{\,\mbox{str}\,}

\def\sTr{\,\mbox{sTr}\,}

\DeclareMathOperator{\cx}{\square}

\def\al{\alpha}
\def\be{\beta}
\def\ch{\chi}
\def\ga{\gamma}
\def\de{\delta}

\def\ep{\epsilon}

\def\ka{\kappa}
\def\la{\lambda}
\def\na{\nabla}
\def\pa{\partial}

\def\si{\sigma}
\def\om{\omega}
\def\ph{\varphi}

\def\Ga{\Gamma}
\def\De{\Delta}

\usepackage{float} %%%   Force figure placement in text
\usepackage{cite}  %%%   Mulitple reference citation
\usepackage{titlesec}

\usepackage{ulem} %%risca o texto selecionando ao meio por linha

\titleformat*{\section}{\large\bfseries}
\titleformat*{\subsection}{\normalsize\bfseries}

\begin{document}

%opening
%%  \title{Conformal anomaly in a vector field model with auxiliary scalar field}
\begin{center}
	
{\Large Trace anomaly for a conformal $2D$ vector field model}
	\vskip 6mm
	
%% \textbf
{\large Samuel W. P. Oliveira} %% $^{a,b}$
%% {Samuel William de Paulo Oliveira}
\footnote{E-mail address: sw.oliveira55@gmail.com}
\quad
and	
\quad
{\large Ilya L. Shapiro} %% $^{b}$
\footnote{
		E-mail address: ilyashapiro2003@ufjf.br}
	%%%%%%%%%%%%%%%%%%%%%%%%%%%%%%%%%
	%% G. Y. Oyadomari, o Samuel gostaria que aparecesse como S. W. P. Oliveira
	%%%%%
\vskip 6mm
	
%% 	   $a)$ \
	PPGCosmo,
	Universidade Federal do Esp\'{\i}rito Santo
	\\
	Vit\'oria, 29075-910, ES, Brazil
	\vskip 3mm
	
%%  	$b)$ \
	Departamento de F\'{\i}sica, \ ICE, \
	Universidade Federal de Juiz de Fora
	\\ Juiz de Fora,  36036-900,  MG,  Brazil

\end{center}
%%%%%%%%%%%%%%%%%%%%%%%%%%%%%%%%
\vskip 5mm

\centerline{\textbf{Abstract}}
%% \vskip 1mm
\begin{quotation}
	
\noindent
The trace anomaly and anomaly-induced action are evaluated for the
two-dimensional $2D$ vector theory with classical conformal symmetry.
Implementing local conformal symmetry while preserving the gauge
invariance requires either giving up locality of the classical action
or, equivalently, introducing an auxiliary scalar field. The
two-dimensional limit in such a theory is singular. However, in the
dimensional regularization, the limit $D \to 2$ in the one-loop
divergence is smooth. As a result, we arrive at the expression for
anomaly, which has a rich general structure, typical for the
dimensions $D \geq 4$. For comparison and completeness, we also
evaluate anomalies for conformal scalar and fermions, also in the
presence of auxiliary external scalars.
\vskip 3mm

\vskip 1mm

%%    	83E99  	None of the above, but in this section
%%	   83C50   %% 	Electromagnetic fields in general relativity and gravitational theory
\end{quotation}

%\RB \ \ means something has to be modified or was changed

%%%%%%%%%%%%%%%%%%%%%%%%%%%%%%%%%%%%%	
%\newpage

%%%%%%%%%%%%%%%%%%%%%%%%%%%%%%%
%%%%%%%%%%%%%%%%%%%%%%%%%%%%%%%	
\section{Introduction}
\label{s1}

\noindent
The anomalous violation of local conformal symmetry
\cite{CapDuf-74,duff77}
was discovered in the $4D$ (four-dimensional) semiclassical gravity.
The trace (conformal or Weyl) anomaly plays a central role in many
theoretical developments and important physical applications (see,
e.g., \cite{duff94}). On the other hand, the anomaly in $2D$
(two-dimensional space) represents an independent area of research. 
One can even say that
the trace anomaly in $2D$ has a very special place in forming the
content of modern high energy physics. One of the reasons is a
great importance of the Weyl anomaly for string theory
\cite{polyakov81,Fradkin85,Callan85}. An important aspect is
the Polyakov action, i.e., the first example of what is nowadays
called anomaly-induced action. The analog of this action in $4D$
theory \cite{rie,frts84} plays an important role in the applications
such as Hawking radiation \cite{ChFu} and Starobinsky inflationary
model \cite{star} (see earlier work \cite{fhh} and also
\cite{StabInstab} for the latest theoretical developments).

For most of the applications, the anomaly-induced action is the
compact and useful form of parameterizing the logarithmic form
factors that emerge in quantum corrections. This parametrization
works well because the leading logarithms are directly related to
the UV divergences and may be reduced to the dependence on the
conformal factor of the metric \cite{OUP}. This feature holds
independently of the dimension of spacetime.

On the other hand,
there are big differences between conformal anomaly in $2D$
and in higher $D \geq 4$ even dimensions. One of these differences
is that the $2D$ case is much simpler. Since the $2D$ metric has
only one physical degree
of freedom, in the pure metric case, the anomaly induced action does
not have an ``integration constant'', i.e., the ambiguous term which
is an undefined conformal invariant functional. Starting from $4D$,
this term is always present, regardless of whether it has no link to the UV limit
of the theory.  Another distinction concerns the vector field model. The
classical action of gauge fields is conformal invariant only in $4D$,
such that the $2D$ anomaly consists of the contributions of the
scalar and spinor fields, which are conformal invariant in the
absence of masses. There is no contribution from vectors in
$D \neq 4$. The last critical difference concerns the existing
classification of the possible terms in anomaly \cite{ddi,DeserSchw}.
This classification can be systematically obtained based on
the theorem proved in \cite{tmf84} for the $D=4$ case. Let us note
that the discussion of the simpler version in \cite{OUP} can be
directly extended to other dimensions. According to this theorem,
in the classically conformal theory, the coefficient of the pole in
the one-loop divergences satisfies the conformal Noether identity.
The terms that obey conformal identity may be of three distinct
types. One of those is the ``legitimate'' conformal invariants
($c$-terms). There are also nonconformal structures satisfying
the Noether identity ($N$-terms). The list of these terms includes
the unique topological term and surface terms. In the even
dimensions $D \geq 4$, there are all three types of terms. On the
contrary, in $2D$ there is only the topological term because
$c$-terms and total derivatives are algebraically impossible. For
this reason, the anomaly-induced action reduces to the Polyakov
action, for both boson (scalar) and fermion fields \cite{polyakov81}.

Concerning the gauge fields, there are several ways to conformally
extend the vector field model to the dimension $D \neq 4$. In
particular, this can be done by giving up the gauge
invariance \cite{DeserNepo}. The result of this work was used and
discussed from different viewpoints in \cite{Osborn} and obtained
in a different way in \cite{ARSW}. Indeed, this version of the
conformal vector model is not appropriate for discussing anomalies
because violating gauge invariance means extra degrees of freedom
and possible loss of unitarity. On the other hand, in \cite{ARSW} 
were proposed a few alternative ways of extending the conformal 
vector model to $D \neq 4$. These approaches imply either giving 
up the locality of the action following an earlier proposal of 
\cite{BoxAno}, or introducing an extra scalar field. At the classical
level, two of the approaches are equivalent, but in all these models,
the limit $D \to 2$ is singular.

In what follows we shall see that, independent from the
classical singularity of the $D \to 2$, one can formulate a quantum
theory of the vector field in the dimension $D = 2 + \ep$, as part
of the dimensional regularization scheme. The unexpected output
is that the unique singularity in the one-loop divergences is the
usual $1/\ep$ pole. As a result, one can consistently evaluate the
trace anomaly in the $2D$ model. This anomaly depends on both
metric and the extra scalar field, which is a remnant of the
construction of conformal vector theory in $D \geq 4$. Different
from the purely metric case mentioned above, in this case, there
are all three possible types of terms in the anomaly. For the sake
of completeness, we also consider the $2D$ one-loop trace anomaly
for self-interacting scalar field and for the fermion with Yukawa
coupling to external scalar.

The paper is organized as follows. In Sec.~\ref{s2}, we derive the
one-loop divergences for scalar and fermion fields in $2D$.
Sec.~\ref{s3} is devoted to the conformal vector field. We show that
 introducing an auxiliary scalar field to provide conformal invariance
 generates scalar-dependent terms in the final expression for the
 one-loop divergences at  $D \to 2$. The conformal anomaly and the
 anomaly-induced effective action are discussed in Sec.~\ref{s4}.
 Finally, in Sec.~\ref{ConcDisc} we draw our conclusions.

%%%%%%%%%%%%%%%%%%%%%%%%%%%%%%%
%%%%%%%%%%%%%%%%%%%%%%%%%%%%%%%
\section{One loop divergences for scalars and fermions}
\label{s2}

\noindent
Using the heat-kernel technique, the one-loop divergences are
related to the first order of the expansion of the evolution operator
for the bilinear form of the action \cite{DeWitt65},
\beq
\hat{H} %% _{hk}
&=&
\cx \,+\, 2\hat{h}^\la\na_\la \,+\, \hat{\Pi}.
\label{Hhk}
\eeq

The coefficient of the expansion that defines the logarithmic
divergences for a two-dimensional theory is given by
\beq
\hat{P}
&=&
\hat{\Pi} +  \dfrac{\hat{1}}{6}R
- \na_\la\hat{h}^\la - \hat{h}_\la\hat{h}^\la.
\label{P}
\eeq
The expression for the divergences at $D \to 2$, in dimensional
regularization, takes the form
\beq
\Ga^{(1)}_{div}
&=&
\dfrac{i}{2}\,\sTr \log \,\hat{H}%% _{hk}
\,=\,-\,\dfrac{\mu^{D-2}}{\ep}\int d^D x\sqrt{|g|} \, \str  \hat{P},
\label{div}
\eeq
where $\ep = 2\pi(D-2)$, $\mu$ is an artificial dimensional
parameter, $\sTr$ and $\str$ are functional and usual matrix
super-traces, taking into account Grassmann parities of
quantum fields.

%%%%%%%%%%%%%%%%%%%%%%%%%%%%%%%%%%%%%%%%%
%%%%%%%%%%%%%%%%%%%%%%%%%%%%%%%%%%%%%%%%%
%%%%%%%%%%%%%%%%%%%%%%%%%%%%%%%%%%%%%%%%%
\subsection{Scalar field with self-interaction}

Consider a scalar field model with an arbitrary interaction term
\beq
S_{sc}
&=&
\int d^D x\sqrt{|g|}\,\,\Big\{\dfrac{1}{2}(\na\ph)^2  - \eta^2 V(\ph)\Big\},
\label{actscal}
\eeq
where we use the condensed notation $(\na\ph)^2= g^{\al\be}
\pa_\al\ph \pa_\be\ph$,  $V(\ph)$ is a potential function, and $\eta$
is an external scalar field. We consider
the $D$-dimensional spacetime for the reason of convenience. The
conformal transformations of metric and two scalars follow the rules
\beq
g_{\mu\nu}=e^{2\si}\,\bar{g}_{\mu\nu},
\qquad
\ph=\bar{\ph}\,e^{\frac{2-D}{2}\si},
\qquad
\eta=e^{-\si}\,\bar{\eta}\,,
\qquad
\si=\si(x).
\label{contransscal}
\eeq
Using the background field method, the bilinear form of the action
(\ref{actscal}) corresponds to the second order of the expansion in
the quantum field $\ch$, where
\beq
\ph \rightarrow \ph' = \ph+\ch,
\label{bfm-scal}
\eeq
with $\ph$ being the background field. The result is
\beq
\hat{H} &=& \cx\,+\,\eta^2V''(\ph),
\eeq
where $V''(\ph)$ is the second derivative of the potential.
The divergences are expressed as
\beq
\Ga^{(1)}_{\rm div\,sc}
&=&
-\,\dfrac{\mu^{D-2}}{\ep}
\int d^D x\sqrt{|g|}\, \Big\{\,\, \dfrac{1}{3}R + 2\eta^2V''(\ph)\,\Big\}.
\label{divsscal}
\eeq
It is easy to see that the integrand of this expression includes the
usual topological term $R$ and also the $c$-term related to $V''(\ph)$.
The last property holds after taking the limit $D\to 2$ in the
integrand, as it is expected in the general framework \cite{tmf84}.

%%%%%%%%%%%%%%%%%%%%%%%%%%%%%%%%%%%%%
%%%%%%%%%%%%%%%%%%%%%%%%%%%%%%%%%%%%%
%%%%%%%%%%%%%%%%%%%%%%%%%%%%%%%%%%%%%
\subsection{Fermions}

The fermionic action of our interest has a background scalar field $\eta(x)$,
\beq
S_{fer}
&=&
\int d^D x \sqrt{|g|}\,\,i\,
\bar{\Psi} \big(\ga^\mu\na_\mu - ih\,\eta\big) \Psi \,,
\label{actferm}
\eeq
where $h$ is the Yukawa coupling. The  conformal transformation rule 
for $\eta$ is the same as for an external scalar in (\ref{contransscal}),
\beq
g_{\mu\nu}=e^{2\si}\,\bar{g}_{\mu\nu},
\qquad
\Psi=e^{\frac{1-D}{2}\si}\,\bar{\Psi},
\qquad
\eta=e^{-\si}\,\bar{\eta}.
\label{confer}
\eeq
The operator of type \eqref{Hhk} can be derived by the doubling
of the Hessian, that leads to
\beq
\hat{H}
&=&
\cx - \dfrac{\hat{1}}{4}\,R - h^2\eta^2\,\hat{1} - 2ih\,\eta\,\ga^\mu\na_\mu
- ih\,\ga^\mu(\na_\mu\eta),
\label{Hferm}
\eeq
and the divergences have the form
\beq
\Ga^{(1)}_{\rm div\,fer}
&=&
-\, \dfrac{\mu^{D-2}}{\ep}
\int d^{D} x\sqrt{|g|}\, \Big\{
-\dfrac{1}{6}\,R+2h^2\,\eta^2 \Big\}.
\label{divsfer}
\eeq
Once again, the integrand of the divergent part of the one-loop
effective action has the topological term $R$ and the $c$-term
$\eta^2$. Remarkably, the invariance of this term is
related to the transformation rule (\ref{confer}). As we already
noted, this rule is different from the one for the scalar field
$\ph$ in (\ref{contransscal}), hence $\ph$ and $\eta$ should
be regarded as different types of a scalar field.

%%%%%%%%%%%%%%%%%%%%%%%%%%%%%%%
%%%%%%%%%%%%%%%%%%%%%%%%%%%%%%%
%%%%%%%%%%%%%%%%%%%%%%%%%%%%%%%	
\section{Vector field divergences}
\label{s3}

Our starting point is the conformal model of an Abelian vector field
constructed in \cite{ARSW}. The invariance is achieved by inserting
an external scalar field $\Phi(x)$ in the classical action
\beq
S_{vec}
&=&
-\,\dfrac{1}{4}\int d^D x \sqrt{|g|}\, \Phi^\tau
F_{\mu\nu}F^{\mu\nu},
\label{actvec}
\eeq
where the constant $\tau=\frac{D-4}{D-2}$ provides the
return to the usual form of the action in the limit $D \to 4$.
On the other hand, it has a singularity in the limit  $D \to 2$.
However, since the calculation for the one-loop divergences
is performed in $D\neq 2$, we can handle this calculation
without problems, taking this limit only at the end of the
calculation. The action (\ref{actvec}) is maintaining invariance
under the following local conformal transformation:
\beq
g_{\mu\nu}=e^{2\si}\,\bar{g}_{\mu\nu}, \quad
A_\mu=\bar{A}_\mu, \quad
\Phi(x)=\bar{\Phi}(x)\,e^{(2-D)\si}.
\label{convec}
\eeq
In what follows, we use the parametrization
\beq
\psi(x)=\Phi^\tau(x),
\label{Phipsi}
\eeq
such that the singularity at $2D$ becomes implicit.
%%%%%%%%%%%%%%%%%%%%%%%%%%%%%%%%%%%%%%
%\subsection{Bilinear form of the action for the vector field}

Since the action \eqref{actvec} is bilinear in quantum field $A_\mu$,
we can skip the use of the background field method.
The Faddeev-Popov procedure requires the gauge fixing action
\beq
S_{gf}
&=&
-\, \dfrac{1}{2}\int d^D x \sqrt{-g}\,
\psi\, \big(\na_\mu A^\mu\big)^2\,.
\label{FP}
\eeq
In this case, the operator related to the bilinear form can be
expressed as
\beq
\hat{H}
&=&
\de^\nu_\mu \cx \,-\, R^\nu_\mu
\,+\, \psi^{-1} \Big[ \de^\nu_\mu (\na^\la\psi)\na_\la
- (\na^\nu\psi)\na_\mu + (\na_\mu \psi)\na^\nu \Big]\,.
\label{Heq}
\eeq
This is a particular example of the form (\ref{Hhk}), making
the calculation of (\ref{P}) a standard routine. Anyway, we shall
present some intermediate formulas.

The overall expression for the divergences is given by
\beq
\Ga^{(1)}_{\rm div}
&=&
\dfrac{i}{2}\, \textrm{Tr log}\, \hat{H}
\, -\,i\,\textrm{Tr log}\, \hat{H}_{gh},
\eeq
where the bilinear form of the action of gauge ghosts is
$\hat{H}_{gh}=\cx$, such that this part of the calculation is
pretty much standard.

In the vector field sector, after a small algebra, we get
\beq
\big[\hat{P}\big]^\nu_\mu
&=&
 - \, R^\nu_\mu
 \,+ \,\dfrac{1}{6}\,\de^\nu_\mu\,R
  \, + \,\dfrac{1}{2}\psi^{-2}\,\de^\nu_\mu\,(\na\psi)^2
\,-\,\dfrac{1}{2}\,\psi^{-1}\,\de^\nu_\mu\,(\cx\psi)
\nn
\\
&&
+  \, \,\,\dfrac{(D-2)}{4}\,\psi^{-2}(\na_\mu\psi)(\na^\nu\psi).
\label{Peq}
\eeq
Taking the trace and summing up the ghost part, the divergences are
\beq
\Ga^{(1)}_{\rm div\,vec}
\,=\,
-\,\dfrac{\mu^{D-2}}{\ep}
\int d^D x \sqrt{|g|} \,\,
\bigg\{\dfrac{(D-8)}{6}R
+ \dfrac{(3D-2)}{4\psi^2}(\na\psi)^2
- \dfrac{D}{2\psi} \cx\psi \bigg\}.
\label{divvec1}
\eeq
In order to verify the conformal invariance of this expression in the
limit $D\to 2$, let us  perform a reparametrization of the scalar field
\beq
\psi
&=&
e^\phi \quad
\Longrightarrow
\quad \psi^{-1}(\na_\la\psi)\,=\,\na_\la\phi,
\quad \mbox{etc.}
\label{repaphi}
\eeq
In terms of the new variable, and simplifying the integrand of
(\ref{divvec1}) by replacing $D \to 2$, we arrive at
\beq
\Ga^{(1)}_{\rm div\,vec}
\,=\,
-\, \dfrac{\mu^{D-2}}{\ep}\int d^2 x \sqrt{|g|}\,
\Big\{- R +2(\na\phi)^2 - \cx\phi \Big\}.
\label{divvec2}
\eeq
Assuming that the scalar field $\phi$ does not transform, i.e.,
$\phi =  \bar{\phi}$, the integrand of the last expression is
conformal invariant. In this case, the
expression (\ref{divvec2}) has a familiar form.
Once again, we meet here more than the unique topological term
$R$, as there are also the $c$-term $(\na\phi)^2$ and the surface
term $\cx\phi$. Thus, in the presence of an extra scalar, all three
types of terms emerge in the UV divergences in the vicinity of
the dimension $2D$.

%%%%%%%%%%%%%%%%%%%%%%%%%%%%%%%%%
%%%%%%%%%%%%%%%%%%%%%%%%%%%%%%%%%
%%%%%%%%%%%%%%%%%%%%%%%%%%%%%%%%%
\section{Conformal anomaly and anomaly-induced effective action}
\label{s4}

\noindent
Let us start the derivation of the conformal anomaly from the vector
field case. The other two examples can be dealt with similarly.
We follow the approach of \cite{duff77} with small simplifications
\cite{PoImpo}.
The renormalized one-loop effective action of vacuum has the form
\beq
\Ga_{R}
&=&
S_{\rm class} \, + \, \Ga^{(1)} \, + \, \De S,
\label{GaR}
\eeq
where $S_{\rm class}$ is the classical action of vacuum. According
to the logic of \cite{tmf84}, at the one-loop level $S_{\rm class}$
may be an algebraic sum of conformal ($c$ and $N$) terms without
violating renormalizability. Furthermore,
$\Ga^{(1)}=\Ga^{(1)}_{fin}+\Ga^{(1)}_{div}$ is the sum of
divergent and finite parts of the one-loop contributions. Finally,
$\De S = - \Ga^{(1)}_{div}$ are local counterterms,
\beq
\De S
&=&
\frac{1}{D-2}
\int_D \sqrt{|g|}\,\,
\big\{ aR + \om (\na\phi)^2 + b \cx\phi\big\},
\label{deS}
\eeq
where we introduced the useful notations
\beq
a = b = - \frac{1}{2\pi},
\quad
\om = \frac{1}{\pi}\,,
\quad
\mbox{and}
\quad
\int_D
\,=\,\mu^{D-2} \int d^Dx\,\,.
\label{intD}
\eeq
According to \cite{duff77} (see also \cite{OUP} for detailed
explanations and further discussion), the combination
$S_{\rm class}+\Ga^{(1)}$ is conformal, so the anomaly can be
obtained as
\beq
\mathcal{T}
&=&
\big< \mathcal{T}^\mu_{\ \mu} \big>
\,=\,
-\,\dfrac{2}{\sqrt{|g|}}\,g_{\mu\nu}
\,\dfrac{\de \Ga_R}{\de g_{\mu\nu}}
\,=\,
- \dfrac{1}{\sqrt{|\bar{g}|}}\dfrac{\de}{\de \si}
\,\De S\big(g_{\mu\nu} = \bar{g}_{\mu\nu} e^{2\si} \big)\bigg|,
\label{anoma}
\eeq
where the vertical bar means the limit
$D \to 2,\,\,\si \to 0,\,\,\bar{g}_{\mu\nu} \to g_{\mu\nu}$.

Taking into account the transformation rules (see, e.g., \cite{Stud})
\beq
&&
\sqrt{|g|} \big(\na\phi\big)^2
\,=\, e^{(D-2)\si}\,\sqrt{|\bar{g}|} \big(\bar{\na}\bar{\phi}\big)^2,
\nn
\\
&&
\sqrt{|g|} \cx \phi
\,=\, e^{(D-2)\si}\,\sqrt{|\bar{g}|}
\big[\bar{\cx} \bar{\phi} + (D-2) \big(\bar{\na}^\la \si \big)
\big(\bar{\na}_\la \bar{\phi} \big)\big],
\nn
\\
&&
\sqrt{|g|} R
\,=\,
e^{(D-2)\si}\,\sqrt{|\bar{g}|}\big[\bar{R} 
- 2(D-1) \bar{\cx} \si
- (D-1)(D-2) (\bar{\na} \si)^2 \big)\big]
\label{phitrans}
\eeq
and using the procedure (\ref{anoma}), we immediately arrive
at\footnote{There may be an ambiguity related to the last term, 
but we do not intend to discuss it here.}
%%%%%%%%%%%%%%%%%%%%%%%%%%%%
\beq
\mathcal{T}
&=& aR + \om (\na\phi)^2 + b \cx\phi\,.
\label{anomavec}
\eeq

To better understand the anomaly (\ref{anomavec}) and
compare it to the $4D$ case, we can derive the anomaly-induced
effective action of vacuum. This implies solving the equation
\beq
-\,\dfrac{2}{\sqrt{|g|}}\,g_{\mu\nu}
\,\dfrac{\de \Ga_{\rm ind}}{\de g_{\mu\nu}}
\,=\, \mathcal{T}\,.
\label{anoindac}
\eeq

It proves useful to rewrite this equation as we did earlier in
(\ref{anoma}), i.e.,
\beq
- \dfrac{1}{\sqrt{|\bar{g}|}} \dfrac{\de}{\de \si}
\,\Ga_{\rm ind}\big(g_{\mu\nu} = \bar{g}_{\mu\nu} e^{2\si} \big)\bigg|
&=&
\mathcal{T}_a + \mathcal{T}_\om + \mathcal{T}_b,
\label{anoinsig}
\eeq
where $\mathcal{T}_{a,\,\om,\,b}$ are the three terms in anomaly
(\ref{anomavec}). Different from Eq.~(\ref{anoma}), this time
vertical bar does not include the limit $D \to 2$ because equation
(\ref{anoinsig}) should be solved in $2D$.

The solution follows the general scheme described in \cite{rie}
and formulated in a very general form in \cite{6d} as part of
integrating anomaly in $6D$. The basic element of this solution
is the last formula of (\ref{phitrans}) in the $2D$ limit, i.e.,
\beq
\sqrt{|g|} R
\,=\,\sqrt{|\bar{g}|}\big[\bar{R} - 2 \bar{\cx} \si\big].
\label{Rtrans2}
\eeq
Let us start from the $c$-term $\mathcal{T}_\om$ and guess
the corresponding term in induced action,
\beq
\Ga_\om
&=&
\ka \iint\limits_{x\,y}  R_x \,\,\Big(\frac{1}{\cx}\Big)_{x,y}\,(\na \phi)_y^2\,,
\quad
\mbox{where}
\quad
\int\limits_x = \int d^2x \sqrt{|g(x)|}\,.
\label{Gaom}
\eeq
Since $\cx$ and $(\na \phi)^2$ are conformally covariant objects,
using (\ref{Rtrans2}) one can easily see that (\ref{Gaom}) is linear
in $\si$ and arrive at the solution \ $\ka = \om/2$. In a similar way,
we find
\beq
\Ga_a
&=&
\frac{a}{4} \iint\limits_{x\,y}  R_x \,\Big(\frac{1}{\cx}\Big)_{x,y}\,R_y\,.
\label{Gaa}
\eeq
According to (\ref{Rtrans2}) the solution for the total
derivative $N$-term \ $\mathcal{T}_b$ \ is the local functional
\beq
\Ga_b
&=&
\frac{b}{2} \int\limits_x \,R \phi.
\label{Gab}
\eeq
The last observation is that including the contributions of scalars
(\ref{divsscal}) and fermions (\ref{divsfer}) can be reduced to the
redefinition of the coefficients $a$, $b$ and introducing generalized
$c$-term. For $N_s$ copies of scalars, $N_f$ copies of fermions
and $N_v$ copies of vectors (\ref{actvec}), we get
\beq
&&
a_t = \frac{1}{2\pi}\Big(\frac13 N_s - \frac16 N_f - N_v \Big),
\label{at}
\\
&&
Z = \frac{1}{\pi}\,\big(V'' N_s + h^2 N_f\big) \eta^2 
+ \frac{\om N_v}{2}(\na \phi)^2.
\label{Z}
\eeq
Using these quantities, the anomaly-induced action can be cast in
the form
\beq
 \Ga_{\rm ind}
 \,=\,
S_c
\,+\,
\frac{a_t}{4} \iint\limits_{x\,y}  R_x \,\Big(\frac{1}{\cx}\Big)_{x,y}\,R_y
\,+\,
\frac12
\iint\limits_{x\,y}  R_x \,\Big(\frac{1}{\cx}\Big)_{x,y}Z_y
\,+\,
\frac{bN_v}{2} \int\limits_x \,R \phi.
\quad
\label{Gaind}
\eeq
The remarkable special feature of this expression is the presence of
an arbitrary conformally invariant functional $S_c$. In the purely
metric background this term is absent, but in the presence of
external scalars $\eta$ and $\phi$, there is an infinite amount of
terms which can constitute $S_c$. For instance, such a term can be
given by a power series
of the terms $\int (\na \phi)^2 \big[\cx^{-1}(\na \phi)^2 \big]^n$.
Thus, the anomaly-induced action in $2D$ may be qualitatively similar
to the analogous constructions in the even dimensions $D \geq 4$.

To complete the anomaly-induced approach, we reformulate the
nonlocal induced action in the local form, by means of two auxiliary
fields, as it is customary in $D \geq 4$ \cite{a,OUP,6d}. One can
note that the Polyakov action needs only one such scalar field and
the first work on the anomaly-induced action in $4D$ \cite{rie} also
used one scalar.

The first step is to rewrite  (\ref{Gaind}) in the symmetric, or
Gaussian form, as
\beq
&&
\Ga_{\rm ind}
 \,=\,
S_c
\,+\,
\frac{a_t}{4} \iint\limits_{x\,y}
\bigg( R + \frac{1}{a_t}Z\bigg)_x
\,\Big(\frac{1}{\cx}\Big)_{x,y}\,
\bigg( R + \frac{1}{a_t}Z\bigg)_y
\nn
\\
&&
\qquad \qquad % \qquad
-\,\,
\frac{1}{4a_t} \iint\limits_{x\,y}
Z_x\,\Big(\frac{1}{\cx}\Big)_{x,y}\,Z_y
\,+\,
\frac{bN_v}{2} \int\limits_x \,R \phi.
\quad
\label{Gaindsym}
\eeq
The explicit form of the local version of the action depends on the
sign of the coefficient $a_t$ in (\ref{at}). Since our main interest
is about the vector field, we arrive at the following local form of
induced action:
\beq
&&
\Ga_{\rm ind\,loc}
 \,=\,
S_c
\,+\, \frac{bN_v}{2} \int\limits_x \,R \phi
\,+\, \int_x\,\bigg\{
\frac12 \chi \cx \chi - \frac12 \psi \cx \psi
\nn
\\
&&
\qquad \qquad % \qquad
- \,\,\sqrt{- \,\frac{a_t}{2}} \chi  \Big(R + \frac{1}{a_t}Z \Big)
\,+ \,\sqrt{- \,\frac{1}{2a_t}} \,\psi Z \bigg\}.
\quad
\label{Gaindloca}
\eeq
For the applications, the boundary and/or initial conditions which
define the Green functions of the operator $\cx$ in (\ref{Gaind}),
are mapped into the corresponding conditions for the new fields $\chi$
and $\psi$ in the local version of the action (\ref{Gaindloca}). The
terms with the field $\psi$ are conformal, but our experience
with the $4D$ applications (see the discussion and further references
in \cite{PoImpo}) shows that it is useful not to include this part in
$S_c$, since only with these terms the induced action provides good
correspondence to the UV leading logarithms.

%%%%%%%%%%%%%%%%%%%%%%%%%%%%%%%%
%%%%%%%%%%%%%%%%%%%%%%%%%%%%%%%%
\section{Conclusions}
\label{ConcDisc}

\noindent
The trace anomaly in a $2D$ space is one
of the best studied subjects in quantum field theory. However, all
the existing results concern contributions of scalars (including
sigma models) and fermions, required by supersymmetry. Until
recently, inclusion of a vector field was not possible because this
field is conformally invariant only in $4D$, but not in $2D$.

In this letter, we use the previously developed model \cite{ARSW}
of conformal
vector field in $D \neq 4$ to explore the vector anomaly in $2D$.
One of the conformal models constructed in this work is especially
simple, as the symmetry does not require violation of gauge
invariance or introducing a nonlocal classical action. Instead,
there is an auxiliary scalar field. In the original formulation of
\cite{ARSW}, the action of this model is singular in the limit
$D \to 2$. However, using dimensional regularization, we can perform
calculations without taking this limit. As a result, one loop quantum
corrections produce a conformal anomaly which is not degenerate
in $2D$. The anomaly depends not only on the metric but also on
the auxiliary scalar field, producing interesting new structures.

The vector model in $2D$ does not have physical degrees
of freedom, different from scalars or fermions. An interesting detail
is that the sign of the vector contribution to the traditional anomaly
(\ref{at}) is the same as for the spinors, opposite to the one for
scalars. On the other hand, the presence of an external scalar
makes the structure of the anomaly richer, similar to the one in the
even dimension $D \geq 4$. As a result, the $2D$ anomaly
possesses all possible terms of the general classification
\cite{DeserSchw}. We have shown that the corresponding
anomaly-induced action gains the full form with two different  Green
functions, or with two auxiliary scalar fields, typical for the
dimension four \cite{a} and beyond \cite{6d}.

%%%%%%%%%%%%%%%%%%%%%%%%%%%%%%%%
%%%%%%%%%%%%%%%%%%%%%%%%%%%%%%%%
\section*{Acknowledgements}
%% \label{secAck}

\noindent
S.W.P.O. is grateful to Coordena\c{c}\~{a}o de Aperfei\c{c}oamento de
Pessoal de N\'{\i}vel Superior -- CAPES (Brazil) for supporting his Ph.D.
project, and to Funda\c{c}\~{a}o de Amparo \`{a} Pesquisa e
Inova\c{c}\~{a}o do Esp\'{\i}rito Santo -- FAPES (Brazil) for supporting
this project. The work of I. Sh. is partially supported by Conselho Nacional de
Desenvolvimento Cient\'{\i}fico e Tecnol\'{o}gico - CNPq under the grant
305122/2023-1.

%%%%%%%%%%%%%%%%%%%%%%%%%%%%%%%%%%%
\vskip 4mm

%%%%%%%%%%%%%%%%%%%%%%%%%%%%%%%%%%%
%%%%%%%%%%%%%%%%%%%%%%%%%%%%%%%%%%%
%%%%%%%%%%%%%%%%%%%%%%%%%%%%%%%%%%%

\end{document}